\documentclass[aps,twocolumn,groupedaddress]{revtex4}
\usepackage{psfig}%,showkeys}
\begin{document}
% You should use BibTeX and apsrev.bst for references
\bibliographystyle{apsrev}

\preprint{condmat/xy}

% Optional argument for running titles on pages
%\title[]{}
\title[Schweitzer/Ebeling/Tilch: Statistical Mechanics of Canonical
...]{Statistical Mechanics of Canonical-Dissipative Systems \\ and
  Applications to Swarm Dynamics} \author{Frank Schweitzer}
\email[]{schweitzer@gmd.de} \homepage[]{http://ais.gmd.de/~frank/}
%\thanks{}
\affiliation{Real World Computing Partnership - Theoretical
    Foundation GMD Laboratory, Schloss Birlinghoven, 53754 Sankt
    Augustin, Germany}
\author{Werner Ebeling}
\email[]{ebeling@physik.hu-berlin.de}
\homepage[]{http://summa.physik.hu-berlin.de/tsd/}
\affiliation{Institute of Physics, Humboldt University,
    Invalidenstra{\ss}e 110, 10115 Berlin, Germany}
\author{Benno Tilch}
\email[]{benno@theo2.physik.uni-stuttgart.de}
\homepage[]{http://www.theo2.physik.uni-stuttgart.de/}
\affiliation{II.  Institute of Theoretical Physics, University of
    Stuttgart, Pfaffenwaldring 57/III, D-70550 Stuttgart, Germany}



\date{15 March 2001}

\begin{abstract}
  We develop the theory of canonical-dissipative systems, based on the
  assumption that both the conservative and the dissipative elements of
  the dynamics are determined by invariants of motion. In this case,
  known solutions for conservative systems can be used for an
  extension of the dynamics, which also includes elements such as the
  take-up/dissipation of energy. This way, a rather complex dynamics can
  be mapped to an analytically tractable model, while still covering
  important features of non-equilibrium systems. 
  
  In our paper, this approach is used to derive a rather general swarm
  model that considers (a) the energetic conditions of swarming, i.e. for
  active motion, (b) interactions between the particles based on global
  couplings. We derive analytical expressions for the non-equilibrium
  velocity distribution and the mean squared displacement of the swarm.
  Further, we investigate the influence of different global couplings on
  the overall behavior of the swarm by
  means of particle-based computer simulations and compare them with the
  analytical estimations.
\end{abstract}
%\maketitle must follow title, authors, abstract and \pacs
\maketitle

\renewcommand{\bbox}[1]{\mbox{\boldmath $#1$}}
\newcommand{\mean}[1]{\left\langle #1 \right\rangle}
\newcommand{\abs}[1]{\left| #1 \right|}
\newcommand{\eqn}[1]{eq. (\ref{#1})}
\newcommand{\Eqn}[1]{Eq. (\ref{#1})}
\newcommand{\sect}[1]{Sect. \ref{#1}}
\newcommand{\eqs}[2]{eqs. (\ref{#1}), (\ref{#2})}
\newcommand{\pic}[1]{Fig. \ref{#1}}
\newcommand{\name}[1]{{\rm #1}}
\newcommand{\bib}[4]{\bibitem{#1} {\rm #2} (#4): #3.}
\frenchspacing

\section{Introduction}

The collective motion of biological entities, like schools of fish,
flocks of birds, herds of hoof animals, or swarms of insects, recently
also attracted the interest of physicists. Here, the question how a
long-range order between the moving entities can be established is of
particular interest. Consequently, some of the more biologically centered
questions of swarming behavior, namely about the resons of swarming or
the group size dependence, have been dropped so far in physical swarm
models. The main focus was rather on the emergence of coherent motion in
a ``swarm'' of locally or globally coupled particles.

For the coupling different assumptions have been proposed (cf. also \sect{4}), such as the coupling of the particles' individual orientation (i.e.
direction of motion) to the mean orientation of the swarm
\citep{czirok-et-96,czirok-vicsek-00}, or the coupling of the particles'
individual position to the mean position (center of mass) of the swarm
\citep{mikhailov-zanette-99}. On the other hand, also some more local
couplings have been considered, such as the coupling of the particle's
individual velocity to a local average velocity
\citep{toner-tu-95,vicsek-et-95,czirok-et-99,czirok-vicsek-00}.  A different class of models
further assumes a local coupling of the particles via a self-consistent
field that has been generated by them
\citep{lsg-mieth-rose-malch-95,lsg-fs-mieth-97}.  This models the case of
chemical communication between the particles as widely found in biology.
For example, the streaming behaviour of myxobacteria
\citep{%stevens-90,
  stevens-95,ben-jacob-et-95-b,hoefer-et-95}, or the directed motion of
ants \citep{calenbuhr-deneubourg-90,edelstein-keshet-94,fs-lao-family-97}
or the coherent movement of cells \citep{schienbein-gruler-95} have been
described and modeled by means this kind of local coupling.

But long-range or short-range coupling of the particles is only one of
the prerequisites that account for swarming. Another one is the
\emph{active motion} of the particles.  Of course, particles can also
move passively, driven by thermal noise, by convection, currents or by
external fields.
This kind of driving force however does not allow the particle to change
its direction of motion, or velocity etc.  \emph{itself}.  Recent models
of \emph{self-driven} particles which are used to simulate swarming
behavior \cite{vicsek-et-95,albano-96,helbing-vicsek-99} usually just
postulate that the entities move with a certain non-zero velocity,
without considering the \emph{energetic} implications of active motion.
In order to do so, we need to consider that the many-particle system is
basically an \emph{open} system which is driven into non-equilibrium.

To this end, our approach to swarming is based on the theory of
\emph{canonical-dissipative systems}.  This theory -- which is not so
well-known even among experts -- results from an extension of the
statistical physics of Hamiltonian systems to a special type of
dissipative systems, the so-called canonical-dissipative systems
\cite{graham-73,haken-73,hongler-ryter-78,eb-engelh-80,eb-81,feistel-eb-89,eb-00-b}. 
The term \emph{dissipative} means here that the system is
non-conservative and the term \emph{canonical} means, that the
dissipative as well as the conservative parts of the dynamics are both
determined by a Hamilton function $H$ (or a larger set of invariants of
motion, see \sect{2} for details).

This special
assumption allows in many cases exact solutions for the distribution
functions of many-particle systems, even in far-from-equilibrium
situations. This was known already to pioneers as Poincare, Andronov and
Bautin 
who gave exact solutions for a special class of nonlinear oscillators
\citep{andronov-witt-chaikin-65}.  In recent work \citep{makarov-eb-velarde-00} the
properties of canonical-dissipative systems were used to find exact
solutions for \emph{dissipative solitons} in Toda lattices which are
pumped with free energy from external sources. The starting point was the
well-known Toda theory of soliton solutions in one-dimensional lattices
with a special nonlinear potential (exponential for compression and
linear for expansion) of the springs \cite{toda-81,toda-83}, which was
then extended towards a non-conservative system
\cite{graham-73,eb-81,feistel-eb-89}.
In another work \cite{eb-00} an application to systems of Fermi and
Bose particles was given.

In this paper, we will apply the theory of canonical-dissipative systems
to the dynamics of swarms. We start with an outline of the general theory
in \sect{2}, describing first the deterministic and then the stochastic
approach. As a first step towards the dynamics of swarms, in \sect{3} we
investigate the \emph{energetic conditions} of swarming, i.e. we discuss
the conditions for active motion and their impact on the distribution
function and the mean squared displacement of the particles. In \sect{4}
we introduce \emph{global interactions} between the particles which may
account for swarming, i.e. for the maintenance of \emph{coherent} motion
of the ensemble.  The global interactions are choosen with respect to the
general theory of canonical-dissipative systems.  By means of computer
simulations we show how different kinds of coupling affect the overall
behavior of the swarm. In \sect{5} we conclude with some ideas how to
generalize the approach presented in this paper.

%%%%%%%%%%%%%%%%%%%%%%%%%%%%%%%%%%%%%%%%%
\section{General Theory of Canonical-Dissipative Systems} 
\label{2}
\subsection{Dynamics of Canonical-Dissipative Systems}
\label{2.1}

Let us consider a mechanical many-particle sytem with $f$ degrees of
freedom $i = 1,...,f $ and with the Hamiltonian
$H(q_1\,\ldots\, q_f,\,  p_1\, \ldots\, p_f)$
The corresponding equations of motion are
\begin{equation}
\label{H-qp}
\frac{d q_i}{d t} = \frac{\partial H}{\partial p_i} \;\;;\quad
\frac{d p_i}{d t} = - \frac{\partial H}{\partial q_i}
\end{equation}
Each solution of the system of equations (\ref{H-qp})
\begin{equation}
\label{pq-t}
p_i = p_i(t)\;\;; \quad q_i = q_i (t)
\end{equation}
defines a trajectory on the plane $H =
E=$const. This trajectory is determined by the initial conditions, 
and also the energy $E = H(t=0)$
is fixed due to the initial conditions. We construct now a
\emph{canonical-dissipative system} with the same Hamiltonian, where
$g(H)$ denotes the \emph{dissipation function}:
\begin{equation}
\label{H-qp-diss}
\frac{d p_i}{d t} = - \frac{\partial H}{\partial q_i} -
           g(H) \frac{\partial H}{\partial p_i}
\end{equation}
In order to elucidate this kind of canonical-dissipative dynamics, we
will consider different examples for the dissipation function in the
following.  In general we will only assume that $g(H)$ is a nondecreasing
function of the Hamiltonian. The canonical-dissipative system,
\eqn{H-qp-diss}, \cite{graham-73,eb-81,feistel-eb-89} does not conserve
the energy because of the following relation:
\begin{equation}
\label{H-t}
\frac{d H}{d t} =  -  g(H) \sum_i \left(
\frac{\partial H}{\partial p_i}\right)^2
\end{equation}
Whether the total energy increases or decreases, consequently depends on
the form of the dissipation function $g(H)$. 
In the simplest case, we may consider a constant friction:
\begin{equation}
\label{gamma0}
g(H) = \gamma_0 > 0
\end{equation}
As far as $g(H)$ is always positive, the energy always decays. 

In a more interesting case the dissipative function $g(H)$ has a root for
a given energy $E_1$: $g(E_1) = 0$. Let us further assume that at least
in certain neighbourhood of $E_1$ the function $g(H)$ is increasing.
With these assumptions the states with $H < E_1$ are pumped with energy
due to the negative dissipation, while energy is extracted from the
states with $H > E_1$. That means, any given initial state with $H(t=0) <
E_1$ will increase its energy until it reaches the shell $H(t) = E_1$,
while any given initial state with $H(t=0) > E_1$ will decrease its
energy until the shell $H(t) = E_1$ is reached, too.  Therefore the
solution of \eqn{H-t} converges to the energy surface $H = E_1$.
On this surface the solution of \eqn{H-qp-diss} agrees
with one of the possible solutions of the original Hamiltonian \eqn{H-qp}
for $H = E_1$.

The simplest ansatz for $g(H)$ with a root $g(H)=0$ is a \emph{linear}
dissipation function:
\begin{equation}
\label{gh-linear}
g(H) = C (H - E_1)
\end{equation}
With respect to \eqn{H-t} and the discussion above, it is obvious for
this case that the
process comes to rest when the shell $H(t) = E_1$ is reached.  The
relaxation time is proportional to the constant $C^{-1}$.

We note that the linear dissipation function \eqn{gh-linear} has found
applications in Toda chains \cite{makarov-eb-velarde-00}.  Here, the fact
that on the shell $H=E_{1}$ the trajectory should obey the original
\emph{conservative} canonical dynamics, has been used to derive exact
solutions for canonical-\emph{dissipative} Toda systems.

A rather general \emph{nonlinear} and non-decreasing dissipation
function, which has been proposed in \cite{eb-00}, reads:
\begin{equation}
\label{gh-general}
g(H) = \gamma_0 - \gamma_1 \frac{(1+A)}{1+A \exp (\beta H)}
\end{equation}
Here $\gamma_0 >0$ represents the normal positive friction, $\gamma_1 >
0$ represents a kind of negative friction.  $A$ is a dimensionless
constant and $\beta$ is a parameter with the meaning of a reciprocal
temperature.  For $\gamma_1 \leq \gamma_0$ the friction is always
positive, i.e. energy is extracted. For the opposite case, $\gamma_1 >
\gamma_0$, we have negative friction and the system is pumped with energy
at least in some parts of the phase space. This allows to drive the
system into situations far from equilibrium, therefore in the following
we may assume $\gamma_{1}>\gamma_{0}>0$. 
  
In the limit $\beta \rightarrow 0$ and $A \rightarrow 0$ the dissipative
function \eqn{gh-general} reduces to the linear case, \eqn{gh-linear},
discussed above. On the other hand, for small values of $\beta$ and
finite $A$ we get
\begin{equation}
\label{gh-med}
g(H) = \gamma_0 -\gamma_{1} \frac{(1+A)}{1+A + A \beta H }
\end{equation}
which yields $g(H) \to (\gamma_0 - \gamma_1) < 0$ for $H\to 0$, and $g(H)
\to \gamma_0$, if $H \to \infty$. This way, the existence of a root
$g(E_1) = 0$ is always guaranteed.

Before investigating a special case for $g(H)$ in \sect{3}, let us
introduce a generalization of the formalism. Instead of a
driving function $g(H)$ which depends only on the Hamiltonian we may
include the dependence on a larger set of invariants of motion, $I_0,
I_1, I_2, ..., I_s$, for example:
\begin{itemize}
\item Hamilton function of the many-particle system:
$I_0 = H$    
\item  total momentum of the many-particle system:
${I}_1 = {P}$
\item total angular momentum of the many-particle system:
${I}_2 = {L}$
\end{itemize}
The dependence on this larger sets of invariants may be considered by
defining  a \emph{dissipative potential} $G(I_0,I_1,I_2,...)$. The
canonical-dissipation equation of motion, \eqn{H-qp-diss} is then
generalized towards:
\begin{equation}
\label{G-qp-diss}
\frac{d p_i}{d t} = - \frac{\partial H}{\partial q_i} -
           \frac{\partial G(I_0,I_1,I_2,...)}{\partial p_i}
\end{equation}
Using this generalized canonical-dissipative formalism, by an appropriate
choice of the dissipative potential $G$ the system may be driven to
particular subspaces of the energy surface, e.g.  the total momentum or
the angular momentum may be prescribed. Different examples for this will
be also discussed in \sect{4}.

%%%%%%%%%%%%%%%%%%%%%%%%%%%%%%%%%%%%%%%%%%%%%%%%%%%%%%%%%
\subsection{Stochastic Theory of Canonical-Dissipative Systems}
\label{2.2}

We will now investigate an
approach to the
stationary probabilities which is based on Langevin and Fokker-Planck
equations. The Langevin equations are obtained by adding a white noise
term $\xi_{i}(t)$ to the deterministic \eqn{H-qp-diss}:
\begin{equation}
\label{langev-gen}
\frac{d p_i}{d t} = - \frac{\partial H}{\partial q_i} -
           g(H) \frac{\partial H}{\partial p_i} + \Big(2 D(H)\Big)^ {1/2}\;\xi_{i}(t)
\end{equation}
The essential assumption is that both the strength of the noise,
expressed in terms of $D(H)$ and the dissipation, expressed in terms of
the friction function $g(H)$ depend only on the Hamiltonian $H$.

With respect to \eqn{langev-gen}, the corresponding Fokker-Planck
equation for the probability distribution $\rho (q_1... q_f, p_1... p_f)$
of the many-particle system reads:
\begin{eqnarray}
\label{fpe-gen}
\frac{\partial \rho}{\partial t} + \sum p_i \frac{\partial \rho}{\partial q_i} - 
\sum \frac{\partial H}{\partial p_i}\frac{\partial \rho}{\partial p_i}
 = \nonumber \\ 
\sum \frac{\partial}{\partial p_i}\left[g(H) \rho + D(H) 
\frac{\partial \rho}{\partial p_i}\right] 
\end{eqnarray}
An exact stationary solution of \eqn{fpe-gen} reads:
\begin{equation}
\label{fpe-0-gen}
\rho^{0} (q_1... q_f,p_1... p_f) = Q^{-1} 
\; \exp \int\limits_0^H d H'\;\frac{g(H')}{D(H')}
\end{equation}
The derivative of $\rho^{0}$ vanishes if $g(H)=0$, which means that the
probability distribution is maximal at the surface $H = E_1$.

For the special case of a linear dissipation function, \eqn{gh-linear},
we find the stationary solution:
\begin{equation}
\label{fpe-0-lin}
\rho^{0} (q_1... q_f,  p_1... p_f) = Q^{-1} \; 
\exp \left(\frac{cH\;(2E_1 - H)}{2D}\right) 
\end{equation}
In the limit of very strong pumping, $H\gg E_{1}$, this probability
distribution reduces to a kind of microcanonical ensemble corresponding
to the energy $E_1$.

The existence of exact solutions for the probability distribution,
\eqn{fpe-0-gen}, allows to derive several thermodynamic functions for the
many-particle system, such as the \emph{mean energy}:
\begin{equation}
\label{mean-U}
U = Q^{-1} \int d H\, J(H)\; H \exp \int\limits_0^H\; 
d H' \;\frac{g(H')}{D(H')}
\end{equation}
where the Jacobian $J(H)$ is defined by:  
\begin{equation}
\label{qp}
 dq_1...d q_f\; d p_1...d p_f = J(H)\; dH
\end{equation}
The \emph{entropy} follows from the Gibbs formula which yields here:
% \begin{widetext}
% % put long equation here
% \begin{equation}
% \label{S}
% S = + k_B \ln Q - k_B \int d H\; J(H) \left[Q^{-1} \exp \int\limits_0^H \;
% \frac{g(H)}{D(H)}\right] \;
% \int\limits_0^H d H'\, \frac{g(H')}{D(H')}
% \end{equation}
%\end{widetext}
\begin{eqnarray}
\label{S}
S = + k_B \ln Q - k_B \int d H\; J(H) \times \nonumber \\ 
\times \left[Q^{-1} \exp \int\limits_0^H \;
\frac{g(H)}{D(H)}\right] \;
\int\limits_0^H d H'\, \frac{g(H')}{D(H')}
\end{eqnarray}
Further, the system has a Lyapunov functional $K$ which is provided by
the Kullback entropy:
\begin{equation}
\label{kullback}
K [\rho,\rho^{0}] = \int dq_1...d q_f\, d p_1...d p_f \; \rho \ln 
\left(\frac{\rho}{\rho^{0}}\right)
\end{equation}
This gives explicitely:
\begin{equation}
\label{kullback2}
K [\rho,\rho_0] = \ln Q - \frac{S}{k_B} + \mean{\int\limits_0^H 
\frac{g(H)}{D(H)}} 
\end{equation}
This functional is always nonincreasing.

If we want to generalize the description by means of the dissipative
potential $G(I_0,I_1,I_2,...)$ as used in \eqn{G-qp-diss}, the situation
is more difficult.  But at least if the noise strength $D(H)$ is a
constant ($D =$const.), the corresponding Fokker-Planck equation
\begin{eqnarray}
\label{fpe-gen-inv}
\frac{\partial \rho}{\partial t} + \sum p_i \frac{\partial \rho}{\partial q_i} - 
\sum \frac{\partial H}{\partial p_i}\frac{\partial \rho}{\partial p_i}
= \nonumber \\ \sum \frac{\partial}{\partial p_i}
\left[\frac{\partial G(I_0,I_1,I_2,...)}{\partial p_i} \rho 
+ D \frac{\partial \rho}{\partial p_i}\right]  
\end{eqnarray}
is still solvable. Due to the invariant character of the $I_k$, the
l.h.s. of \eqn{fpe-gen-inv} disappears for all functions of $I_k$.
Therefore we have to search only for a function $G(I_0,I_1,I_2,...)$, for
which also the collision term disappears.  This way we find the
stationary solution of \eqn{fpe-gen-inv}:
\begin{equation}
\label{fpe-0-gen-inv}
\rho^{0} (q_1... q_f,p_1... p_f) = Q^{-1} 
\; \exp \left(- \frac{G(I_0,I_1,I_2,...)}{D}\right)
\end{equation}
The derivative of $\rho_0$ vanishes if $G(I_0,I_1,I_2,..)= {\rm min.}$
which means that the probability is maximal on the attractor of the
dissipative motion.

%%%%%%%%%%%%%%%%%%%%%%%%%%%%%%%%%%%%%%%%%%%%%%%%%%%%
\section{Energetic Conditions of Swarming}
%Applications of the theory to swarm dynamics}
\label{3}

In this section, we want to apply the general description outlined above
to swarm dynamics. Basically, at a certain level of abstraction a swarm
can viewed as a many-particle system with some additonal coupling which
would account for the typical correlated motion of the entities. In
addition, also some energetic conditions must be satisfied in order to
keep the swarm moving. Thus, \emph{active motion} of the particles with a
non-zero velocity is another basic ingredient for swarming. Since the
conditions for active motion are dropped in many of the currently
discussed swarm models \cite{albano-96,czirok-et-99,vicsek-et-95}, we
first want to investigate the energetic conditions of swarming, before
turning to the second ingredient, coupling of individual motion.

%%%%%%%%%%%%%%%%%%%%%%%%%%%%%%%%%%%%%%%
\subsection{Conditions for Active Motion}
\label{3.1}
If we omit interactions between the particles, the Hamiltonian of the
many-particle system is of the simple form:
\begin{equation}
  \label{Hamil0}
H = \sum_{i=1}^N H_{i} = \sum_{i=1}^N \frac{p_i^2}{2m}   
\end{equation}
This allows to reduce the description level from the $N$-particle
distribution function to the one-particle distribution function:
\begin{equation}
  \label{one-part}
\rho(q_1... q_f,   p_1... p_f) = \prod_{i=1}^{N}
\rho(\bbox{r_{i}},\bbox{p_{i}})   
\end{equation}
where the variable $\bbox{r}_{i}$ is now used for the space coordinate of
the particle $i$, and $\bbox{p}_{i}$ stands for the momentum or, if $m=1$
is used in the following, for the \emph{velocity} of the particle,
respectively. The center of mass of the swarm is defined as
\begin{equation}
  \label{mean-r}
  \bbox{R}=\frac{1}{N}\sum \bbox{r}_{i}
\end{equation}
whereas the mean momentum $\bbox{P}$ and the mean angular momentum
$\bbox{L}$ are defined as: 
\begin{eqnarray}
    \label{mean-p}
%\bbox{I}_1 &=& 
\bbox{P}&=&\frac{1}{N}\sum \bbox{p}_{i}    \\
    \label{mean-l}
%\bbox{I}_2 &=& 
\bbox{L}&=&\frac{1}{N}\sum \bbox{L}_{i} = \frac{1}{N}\sum
\bbox{r}_{i} \times \bbox{p}_{i}    
\end{eqnarray}
With $g(H)=g(p^{2}_{i})$ and $D={\rm const.}$, the Langevin
\eqn{langev-gen} for each particle $i$ reads
% with $m=1$ and 
in the absence of an external potential:% \cite{x}:
\begin{equation}
  \label{langev-pump}
\dot{\bbox{r}_{i}}=\bbox{p}_{i}\;;\;\; 
\dot{\bbox{p}}_{i}= -g(p_{i}^{2})\, \bbox{p}_{i} 
%+\nabla U(\bbox{r})
+ (2\, D)^{1/2} \bbox{\xi}_{i}(t)
\end{equation}
where the strength $D$ of the stochastic force results from the Einstein
relation:
\begin{equation}
\label{fdt}
D = \gamma_0 k_{B} T
\end{equation}
$T$ is the temperature and $k_{B}$ is the \name{Boltzmann} constant.

For the dissipation function $g(p^{2})$, we use again the general
nonlinear ansatz of \eqn{gh-general} in the limit of small values of
$\beta$ and finite $A$, \eqn{gh-med}. By means of the substitutions
\begin{equation}
\label{subst}
(1+A) = c\;;\quad \gamma_{1} (1 + A) = d_{2} s_{0}\; ; \quad 
\beta A = 2 d_{2} 
\end{equation}   
and with $H$ of \eqn{Hamil0}, \eqn{gh-med} can be written in the form:
\begin{equation}
\label{gh-qdv}
g(p^{2}) = \gamma_0 - \frac{s_{0} d_{2}}{c +d_{2} p^{2}}
\end{equation}
We note that all 
noninteracting 
systems with $g=g(p^2)$ are of canonical-dissipative type. 

\Eqn{gh-qdv} agrees with the velocity dependent non-linear friction
function previously used  in a model of active
Brownian particles \cite{fs-eb-tilch-98-let,eb-fs-tilch-99}.
These are driven Brownian particles, which move due to the influence of a
stochastic force, but additionally are pumped with energy due to a
velocity-dependent dissipation function $g(p^{2})$, \eqn{gh-qdv} which is
plotted in \pic{gx-damp}.
\begin{figure}[htbp]
\centerline{\psfig{figure=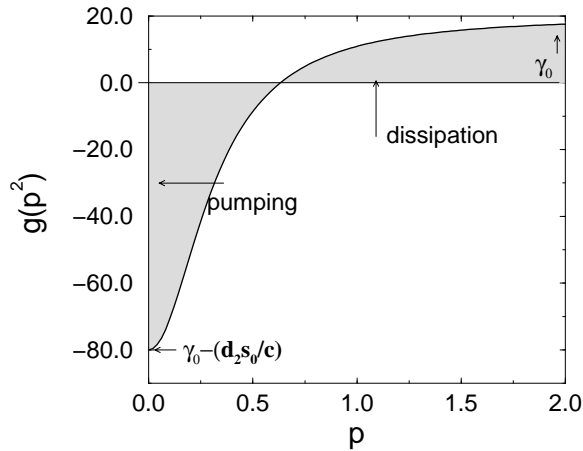,width=7.5cm}}
\vspace*{-3mm}
\caption[]{\label{gx-damp}
  Velocity-dependent dissipation function, $g(p^{2})$, \eqn{gh-qdv} vs.
  $p$. The velocity ranges for ``pumping'' ($g(p^{2})<0$) and
  ``dissipation'' ($g(p^{2})>0$) are indicated.  Parameters: $s_0=10.0$,
  $\gamma_{0}=20.0$, $c=1.0$,
  $d_{2}=10.0$.} 
\end{figure}

The second term of the r.h.s. in \eqn{gh-qdv} which results from the
pumping of energy, has been physically substantiated %derived
in our earlier work \cite{fs-eb-tilch-98-let,eb-fs-tilch-99}.  We have assumed that the Brownian
particles are able to take up energy from the environment at a constant
rate $s_{0}$, which can be stored in an \emph{internal depot}, $e$.  The
internal energy can be converted into kinetic energy with a velocity
dependent rate $d(\bbox{p})=d_{2}\,p^{2}$, which results in an additional
acceleration of the Brownian particle in the direction of movement.
The value of the internal energy depot may be further decreased due to
internal dissipation processes, described by the constant $c$. Thus the
resulting balance equation for the energy depot reads:
\begin{equation}
  \label{depot-t}
  \frac{de}{dt}=s_{0}-c\,e(t)-d_{2}\,p^{2}\,e(t)
\end{equation}
If we assume that the internal energy depot relaxes fast
compared to the motion of the particle, we find the quasistationary
value:
\begin{equation}
  \label{depot-0}
  e_{0}=\frac{s_{0}}{c+d_{2}p^{2}}
\end{equation}
which eventually leads to the second term of the r.h.s. in \eqn{gh-qdv}.

Dependent on the parameters $\gamma_{0}$, $d_{2}$, $s_{0}$, $c$, the
dissipation function, \eqn{gh-qdv}, may have a zero, where the friction is
just compensated by the energy supply. It reads in the considered case:
\begin{equation}
  \label{v-0}
  p_0^2=v_{0}^{2}=\frac{s_0}{\gamma_0} - \frac{c}{d_2}
\end{equation}
We see that for $p<p_{0}$, i.e. in the range of small momentums pumping
due to negative friction occurs, as an additional source of energy for
the Brownian particle. Hence, slow particles are accelerated, while the
motion of fast particles is damped. 

For $s_{0}d_{2}<\gamma_{0}c$, we find no real-valued root of \eqn{v-0}.
This is the case of subcritical pumping, where the particle will move
more or less like a simple Brownian particle. However, provided the
existence of a non-zero momentum $p_{0}$, i.e. for a supercritical
pumping, the particle will be able to move in a ``high velocity'' or
\emph{active} mode
\cite{tilch-fs-eb-99,fs-tilch-eb-00},
which displays several non-trivial features of motion, as will be shown
by means of computer simulations in the next section.

%%%%%%%%%%%%%%%%%%%%%%%%%%%%%%%%%%%%%%a
\subsection{Distribution Function and Dissipative Potential}
\label{3.2}

Due to the pumping mechanism discussed above, the conservation of energy
clearly does not hold for the particle, i.e. we now have a
non-equilibrium, canonical-dissipative system as discussed in \sect{2}.
This results in deviations from the known \name{Maxwellian} velocity
distribution of an equilibrium canonical system. 

As pointed out in \sect{2.2}, the probability density for the velocity
$\rho(\bbox{p},t)$ obeys the Fokker-Planck \eqn{fpe-gen}, which reads for
the special case of the dissipation function,
\eqn{gh-qdv}, and in the absence of an external potential: %, i.e. 
%$U(r_{1},r_{2})\equiv 0$: 
\begin{equation}
  \label{fpe-vt}
  \frac{\partial \rho(\bbox{p},t)}{\partial t}= 
\frac{\partial }{\partial \bbox{p}} \left[ \left(\gamma_{0} -
    \,\frac{d_{2}s_{0}}{c+d_{2}\,p^{2}}\right)\,\bbox{p}\, 
\rho(\bbox{p},t) + D\, \frac{\partial \rho(\bbox{p},t)}{\partial \bbox{p}}
\right]
\end{equation}
We mention that Fokker-Planck equations with nonlinear friction functions
are discussed in detail in \citep{klimontovich-95}.

The stationary solution of \eqn{fpe-vt} is given by \eqn{fpe-0-gen-inv},
which reads in the considered case explicitely:
\begin{eqnarray}
  \label{p0-g}
  \rho^{0}(\bbox{p}) &=& C_{0}\, \exp{\left( - \frac{G_0(p^2)}{D}\right)}
%  \nonumber 
\\ \label{p0-v}
&=& C_{0}\,\left(1+\frac{d_{2}p^2}{c}\right)
  ^{\frac{s_{0}}{2D}}\;
  \exp{\left( - \frac{\gamma_{0}}{2D} \,p^{2}\right)} 
\end{eqnarray}
where $C_{0}$ results from the normalization condition. $G_0(p^2)$ is the
special form of the dissipation potential $G(I_0,I_1,I_2,...)$
considering only $I_{0}=H$ as the invariant of motion, and further $H$ as
given by \eqn{Hamil0}. It reads explicitely:
\begin{equation}
  \label{g-0}
G(I_{0}) = G_0(p^2)=\gamma_{0} \frac{p^2}{2} - \frac{s_{0}}{2} \ln 
\left(1 + \frac{d_{2}p^2}{c}\right) 
\end{equation}
Compared to the Max\-well\-ian velocity distribution of ``simple''
Brownian particles, a new prefactor appears now in \eqn{p0-v} which
results from the additional pumping of energy.  For a subcritical
pumping, \mbox{$s_{0}d_{2}< c \gamma_{0}$}, where we do not find a
real-valued root of the dissipation function, \eqn{gh-qdv}, only an
\emph{unimodal velocity distribution} results, centered around the
maximum $\bbox{p}_0=0$.  However, for supercritical pumping,
\mbox{$s_{0}d_{2}> c \gamma_{0}$}, if the root of $\gamma(p^{2})$ is
real, we find a \emph{crater-like velocity distribution}, which indicates
strong deviations from the Maxwell distribution \cite{erdmann-et-00}.

This is also shown in \pic{pv-1-10} which presents computer simulations
of the velocity distribution of 10.000 particles after a sufficiently
long time (only the x-dimension of the 2-d simulation is shown). For the
supercritical case, two distinct peaks of the velocity distribution are
found at $p_{x}=\{-0.63, +0.63\}$. The values of these maxima agree with
the deterministic result for the stationary velocity, \eqn{v-0}.

We note that non-Maxwellian velocity distributions for active motion have
been also observed experimentally in cells, such as granulocytes
\cite{franke-gruler-90,schienb-gruler-93}.
\begin{figure}[htbp]
\centerline{\psfig{figure=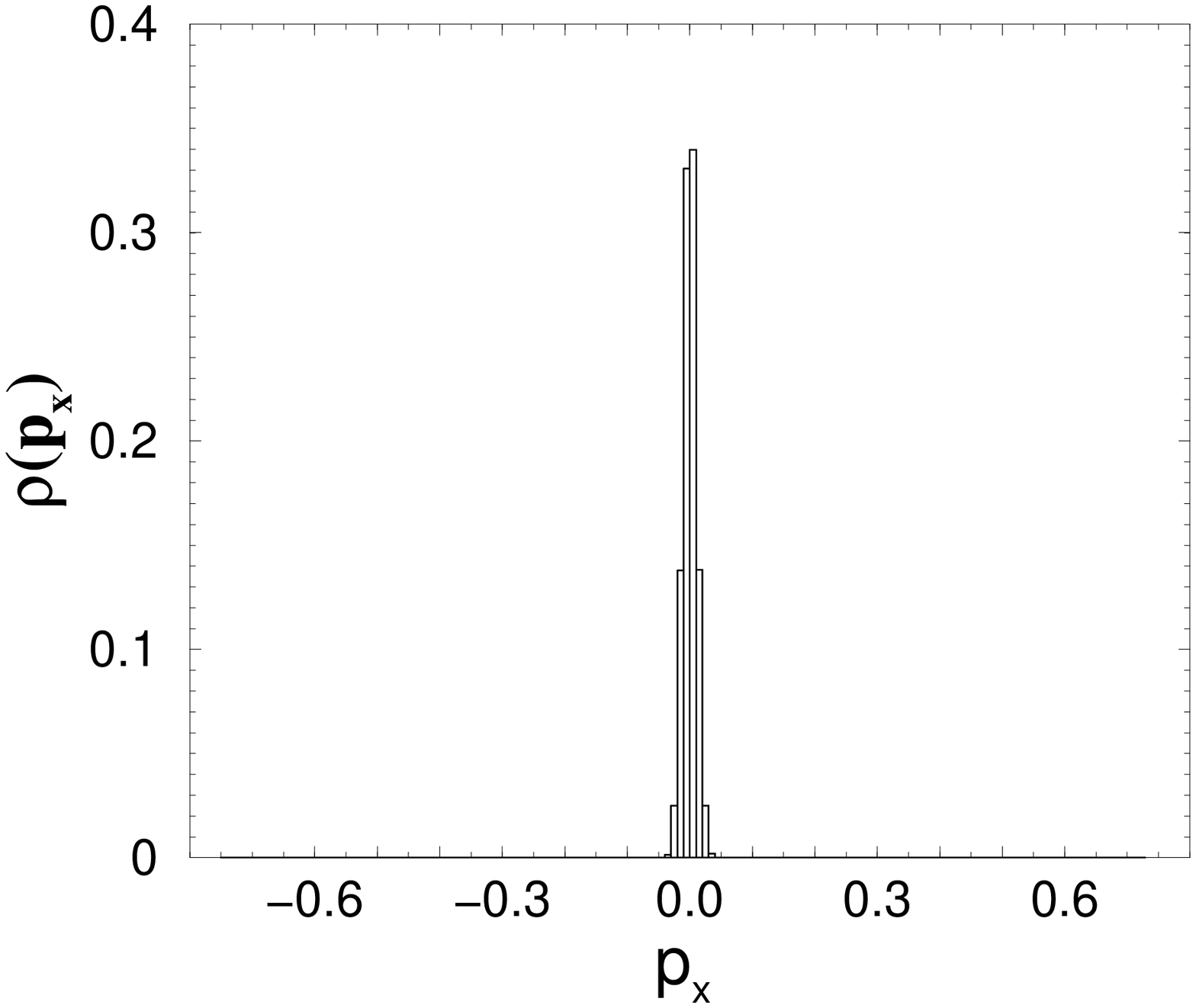,width=7.5cm,angle=-90}}

\centerline{\psfig{figure=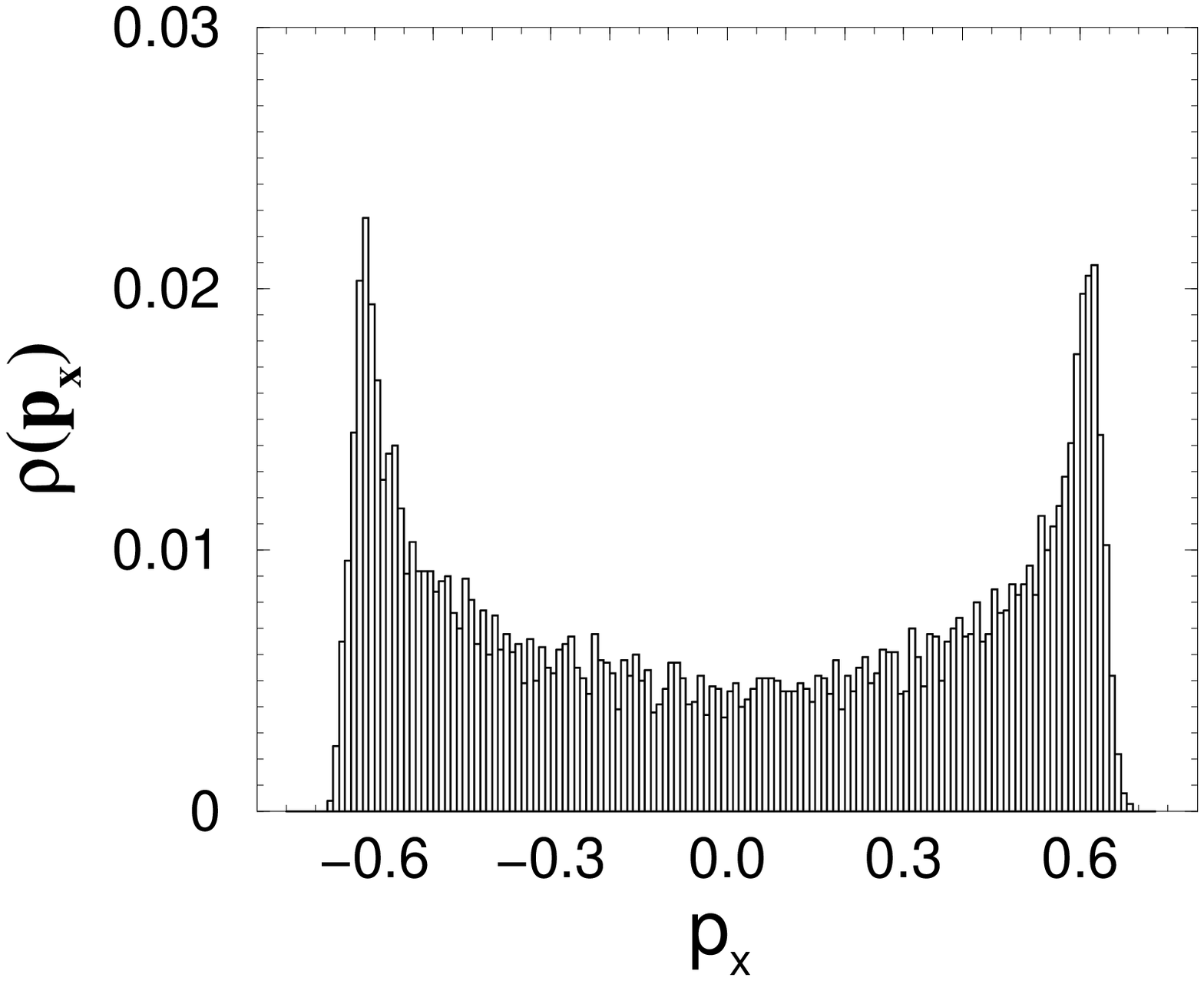,width=7.5cm}}
\vspace*{-3mm}
\caption[]{\label{pv-1-10} 
  Velocity distribution $\rho(p_{x})$ for a swarm of $10.000$ particles
  at $t=1.000$ (i.e. in the stationary regime). Only $p_{x}$ of the 2-d
  simulation is shown. (top) subcritical pumping $d_{2}=1.0$, (bottom)
  supercritical pumping, $d_{2}=10.0$. Other parameters: $D=10^{-3}$,
  $s_{0}=10.0$, $\gamma_{0}=20.0$, $c=1.0$. Initial conditions:
  $\bbox{r}_{i}(0)=\{0.0,0.0\}$, $\bbox{p}_{i}(0)=\{0.0,0.0\}$ for all
  particles.} 
\end{figure}

%%%%%%%%%%%%%%%%%%%%%%%%%%%%%%%%
\subsection{Mean Squared Displacement and Stationary Values}
\label{3.3}

As \pic{pv-1-10} shows,
the momentum distribution is centered around $\bbox{p}=0$ both for
subcritical and supercritical pumping. If we consider a nearly spherical
swarm of particles in the two-dimensional space
\mbox{$\bbox{r}=\{x,y\}$} as in the computer simulations in this
section, its center of mass, \eqn{mean-r} and mean momentum, \eqn{mean-p}
will come to rest. Thus they are not affected by the pumping, but other
quantities are, such as the mean squared displacement:
\begin{equation}
  \label{mean-r2}
  \Delta R^{2}(t)=\mean{\left(\frac{1}{N}\sum \Big[\bbox{r}_{i}(t) -
  \bbox{r}_{i}(0)\Big] \right)^{2}}
\end{equation}
In the limit of pure Brownian motion, it is known that the mean squared
displacement increases in time as:
\begin{equation}
  \label{mean-brown}
  \Delta R^{2}(t)=4 D_{r} t
\end{equation}
where $D_{r}=k_{B}T/\gamma_{0}=D/\gamma_{0}^{2}$ is the spatial diffusion
coefficient.  Thus, \eqn{mean-brown} will be the lower limit for
subcritical pumping of the particles. Contrary, in the case of
supercritical pumping it has been shown \cite{mikhailov-meinkoehn-97,erdmann-et-00}
that the mean squared displacement will grow in time approximately as
\begin{equation}
\label{mean-rpump}
\Delta R^{2}(t) = \frac{2 v^4_0}{D} \,t
\end{equation}
where $v_{0}$ is given by \eqn{v-0}. Consequently, the diffusion
coefficient $D_{r}$ in \eqn{mean-brown} has for the case of supercritical
pumping to be replaced by an \emph{effective} spatial diffusion
coefficient:
\begin{equation}
  \label{d-eff}
  D_{r}^{\rm eff}= \frac{v_{0}^{4}}{2D}=\frac{1}{2D}\left(\frac{s_0}{\gamma_0} - \frac{c}{d_2}\right)^{2}.
\end{equation}
This result holds for noninteracting particles in the limit of
relatively weak noise intensity $D$ and/or strong pumping and will
therefore give an upper limit for $\Delta R^{2}(t)$. We note the high
sensitivity with respect to noise expressed in the scaling with $(1/D)$.

\pic{mean-rr} shows the mean squared displacement of a swarm of 2.000
particles both for the case of sub- and supercritical pumping together
with the theoretical results of \eqs{mean-brown}{mean-rpump}. We see that
for long times the computer simulations for supercritical pumping agree
very well with \eqn{mean-rpump}.
\begin{figure}[htbp]
\centerline{\psfig{figure=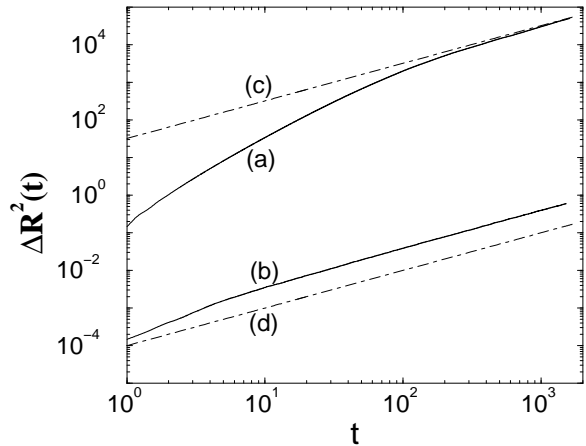,width=7.5cm}}
\caption[]{\label{mean-rr}
  Mean squared displacement $\Delta R^{2}(t)$, \eqn{mean-r2} of a swarm
  of $2.000$ particles as a function of time. (a) supercritical pumping,
  $d_{2}=10.0$, (b) subcritical pumping, $d_{2}=1.0$. $D=10^{-2}$, for
  the other parameters and the initial conditions see \pic{pv-1-10}. The
  additional curves give the theoretical results of \eqn{mean-rpump} (c)
  (upper limit) and \eqn{mean-brown} (d) (lower limit).}
\end{figure}

Another quantity affected by the sub/supercritical pumping is the
stationary velocity $p^{2}_0=v^{2}_{0}$, \eqn{v-0}.
In 2d, these stationary velocities define a cylinder, \mbox{$p_x^2 +
  p_y^2 = p_0^2$}, in the four-dimensional state space
\mbox{$\{x,y,p_{x},p_{y}\}$} which attracts all deterministic
trajectories of the dynamic system \cite{erdmann-et-00}.  \pic{mean-ve}
shows the results of computer simulations for $p^{2}(t)$ for the case of
supercritical pumping. The convergence toward the theoretical result,
\eqn{v-0} can be clearly observed. 
\begin{figure}[htbp]
\centerline{\psfig{figure=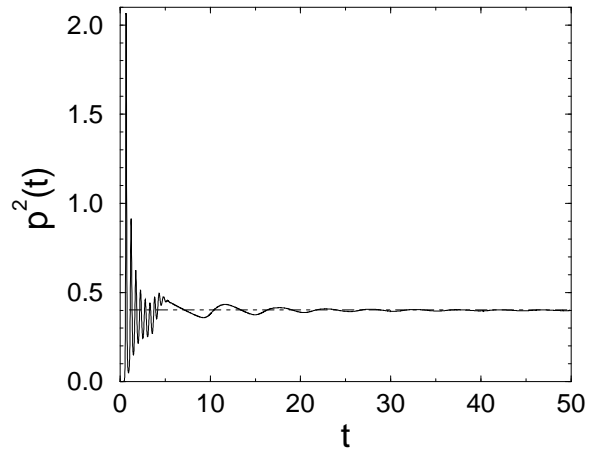,width=7.5cm}}
\caption[]{\label{mean-ve}
  Averaged squared velocity $p^{2}(t)=1/N\sum p_{i}^{2}(t)$,
  \eqn{langev-pump}, of a swarm of $2.000$ particles as a function of
  time.
  $D=10^{-4}$, for the other parameters and the initial conditions see
  \pic{pv-1-10}.  The dash-dotted line gives the stationary velocity,
  \eqn{v-0}.  }
\end{figure}

%%%%%%%%%%%%%%%%%%%%%%%%%%%%%%%%%%%%%%%%%%%%%%%%%%%%
\section{Globally Coupled Swarms}
\label{4}

So far we have neglected any coupling within the many-particle ensemble.
This leads to the effect that the swarm eventually disperses in
the course of time, whereas a ``real'' swarm would maintain its coherent
motion. A common way to introduce correlations between the moving
particles in physical swarm models is the coupling to a mean value. For
example, in \citep{czirok-et-96,czirok-vicsek-00} the coupling of the
particles' individual \emph{orientation} (i.e. direction of motion) to
the mean orientation of the swarm is discussed. 
%Additionally, also a short-range ``hard-core'' repulsion is considered. 
Other versions assume the coupling of the particles' velocity to a
\emph{local average velocity}, which is calculated over a space interval
around the particle \cite{czirok-et-99,czirok-vicsek-00}.

\subsection{Coupling to the Center of Mass}
\label{4.1}

In this paper, we are mainly interested in \emph{global couplings} of the
swarm, which fit it into the theory of canonical-dissipative systems
outlined in \sect{2}.  As the most simple case we may first discuss the
global coupling of the swarm to the \emph{center of mass}, \eqn{mean-r}.
That means the particle's position $\bbox{r}_{i}$ is related to the mean
position of the swarm $\bbox{R}$ via a potential
$U(\bbox{r}_{i},\bbox{R})$. For simplicity, we may assume a parabolic
potential:
\begin{equation}
  \label{parab-int}
  U(\bbox{r}_{i},\bbox{R})= \frac{a}{2}\left(\bbox{r}-\bbox{R}\right)^{2}
\end{equation}
The harmonic potential generates a force directed to the center of mass
which can be used to control the dispersion of the swarm. It reads in the
considered case: 
\begin{equation}
  \label{force}
  \bbox{\nabla} U(\bbox{r})=a \left(\bbox{r}_{i}-\bbox{R}\right)
=\frac{a}{N} \sum_{j=1}^{N}\left(\bbox{r}_{i}-\bbox{r}_{j}\right)
\end{equation}
With \eqn{force}, the corresponding Langevin \eqn{langev-pump} of the
many-particle system reads explicitely:
\begin{equation}
\dot{\bbox{p}}_{i} 
= -g(p_{i}^{2})\, \bbox{p}_{i} 
- \frac{a}{N} \sum_{j=1}^{N}\left(\bbox{r}_{i}-\bbox{r}_{j}\right)
+ (2\, D)^{1/2} \bbox{\xi}_{i}(t)
  \label{2d-new}
\end{equation}
Hence, in addition to the dissipation function there is now an attractive
force between each two particles $i$ and $j$ which depends linearly on
the distance between them. With respect to the harmonic interaction
potential \eqn{parab-int}, we call such a swarm a \emph{harmonic} swarm
\cite{eb-fs-01-pas}.

Strictly speaking, the dynamical system of \eqn{2d-new} is not a
canonical-dissipative one, but as shown in \citep{eb-fs-tilch-99} it may
be reduced to this type by some approximations, which will be also
discussed below.  We note that this kind of swarm model has been
previously investigated in \cite{mikhailov-zanette-99} for the
one-dimensional case, however with a different dissipation function
$g(p^{2})$, for which we use \eqn{gh-qdv} again. Obviously, as shown in
\sect{3.2} swarming will occur only for supercritical conditions.

With the assumed coupling to the center of mass, $\bbox{R}$, the motion
of the swarm can be considered as a superposition of two motions: (i) the
motion of the center of mass itself, and (ii) the motion of the particles
relative to the center of mass.  Taking into account that the noise
acting on the different particles is not correlated, the center of
mass for the assumed coupling obeys a force-free motion, 
\begin{equation}
  \label{langev-RP}
\dot{\bbox{R}}=\bbox{P}\;;\;\; 
\dot{\bbox{P}}= -\frac{1}{N}\;\sum_{i=1}^{N} g(p_{i}^{2})\, \bbox{p}_{i} 
\end{equation}
Because of the nonlinearities in the dissipation function $g(p^{2})$ both
motions (i) and (ii) cannot be simply separated. The term $g(p^{2})$
vanishes only for two cases: the trivial one which is free motion without
dissipation/pumping, or the case of supercritical pumping where
$p_{i}^{2}=p_{0}^{2}$, \eqn{v-0} for each particle. Then, the mean
momentum becomes an invariant of motion, $\bbox{P}(t)=\bbox{P}_{0}={\rm
  const.}$ and the center of mass moves according to
$\bbox{R}(t)=\bbox{R}(0)+\bbox{P}_{0}(t)$. This behavior may also
critically depend on the initial conditions of the particles,
$\bbox{p}_{i}(0)$, and shall be investigated in more detail now.

In \cite{mikhailov-zanette-99} an approximation for the mean velocity
$\bbox{P}(t)$ of the swarm in one dimension is discussed which shows the
existence of two different asymptotic solutions dependent on the noise
intensity $D$ and the initial momentum $\bbox{p}_{i}(0)$ of the
particles. Below a critical noise intensity $D_{c}$, the initial
condition $p_{i}^{2}(0)> p_{0}^{2}$ leads to a swarm the center of which
travels with a constant non-trivial mean velocity, while for the initial
condition $p_{i}^{2}(0)< p_{0}^{2}$ the center of the swarm is at rest.

We can confirm these findings by means of two-dimensional computer
simulations presented in \pic{rr-par8} and \pic{vv-par8}, which show the
mean squared displacement, the average squared velocity of the swarm and
the squared mean velocity of the center of mass for the two different
initial conditions.
\begin{figure}[htbp]
\centerline{\psfig{figure=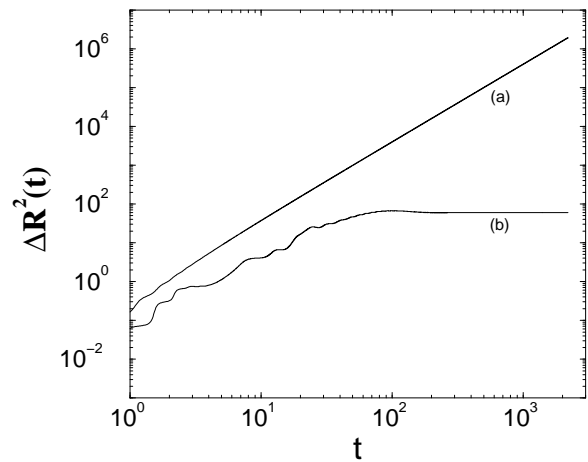,width=7.5cm}}
\caption[]{\label{rr-par8}
  Mean squared displacement $\Delta R^{2}(t)$, \eqn{mean-r2} of a swarm
  of $2.000$ particles coupled to the center of mass. Initial conditions:
  $\bbox{r}_{i}(0)=\{0.0,0.0\}$, (a) $\bbox{p}_{i}(0)=\{1.0,1.0\}$, (b)
  $\bbox{p}_{i}(0)=\{0.0,0.0\}$, for all particles. Parameters: $a=1$,
  $D=10^{-8}$, $d_{2}=10.0$, $s_{0}=10.0$, $\gamma_{0}=20.0$, $c=1.0$.}
\end{figure}
\begin{figure}[htbp]
\centerline{\psfig{figure=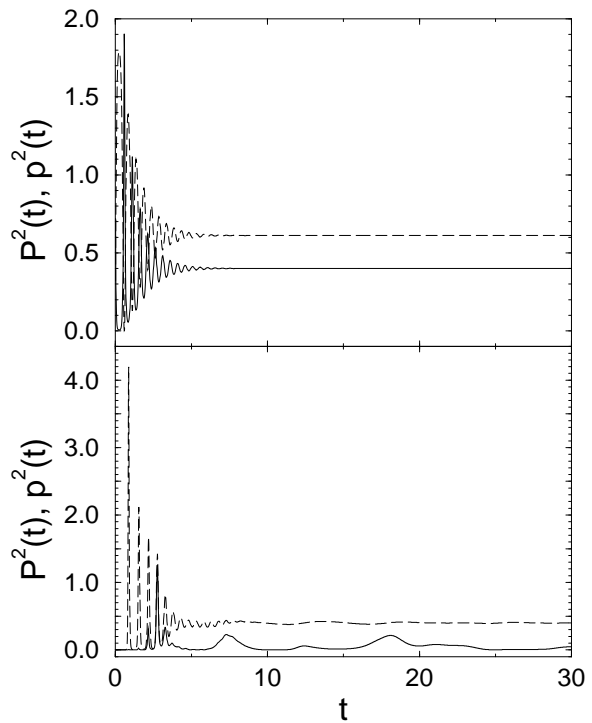,width=7.5cm}}
\caption[]{\label{vv-par8}
  Squared velocity of the center of mass, $P^{2}(t)=\Big(N^{-1}
  \sum_{i}\bbox{p}_{i}(t)\Big)^{2}$ (solid lines) and averaged squared
  velocity $p^{2}(t)=N^{-1}\sum p_{i}^{2}(t)$ (dashed lines) for the
  simulations shown in \pic{rr-par8}: (top) initial conditions (a),
  (bottom) initial conditions (b).}
\end{figure}

For (a) $p_{i}^{2}(0)> p_{0}^{2}$ we find a continuous increase of
$\Delta R^{2}(t)$ (\pic{rr-par8} a), while the velocity of the center of mass reaches a
constant value: $P^{2}(t)=\Big(N^{-1} \sum_{i}\bbox{p}_{i}(t)\Big)^{2}
\to p_{0}^{2}$, known as the stationary velocity of the force-free case,
\eqn{v-0}. The average squared velocity reaches a constant non-trivial
value, too, which depends on the noise intensity and the initial
conditions, $p_{i}^{2}(0)> p_{0}^{2}$, i.e. on the energy initially
supplied (cf \pic{vv-par8} top).

For (b) $p_{i}^{2}(0)< p_{0}^{2}$ we find that the mean squared
displacement after a transient time reaches a constant value, i.e. the
center of mass comes to rest (\pic{rr-par8} b), which corresponds to
$P^{2}(t)\to 0$ in \pic{vv-par8} (bottom). In this case, however, the
averaged squared velocity of the swarm reaches the known stationary
velocity, $p^{2}(t)=1/N\sum p_{i}^{2}(t)\to p_{0}^{2}$.  Consequently, in
this case the energy provided by the pumping goes into the motion
\emph{relative to the center of mass} while the motion of the center of
mass is damped out (cf. also \citep{eb-fs-01-pas}).  Thus, in the
following we want to investigate the relative motion of the particles in
more detail.

Using relative coordinates, $\{x_{i},y_{i}\}\equiv
\bbox{r}_{i}-\bbox{R}$, the dynamics of each particle in the
two-dimensional space is described by four coupled first-order
differential equations:
\begin{eqnarray}
     \dot{x}_{i}  &=&  p_{xi} - P_{x} \nonumber\\    
    \dot{p}_{xi} - \dot{P}_{x} &=&  - g\left(p_{i}^{2}\right) p_{xi} 
- a\, x_{i} + (2\, D)^{1/2}    \xi_{i}(t) \nonumber\\      
    \dot{y}_{i} &=&  p_{yi} - P_{y} \nonumber \\    
    \dot{p}_{yi} -\dot{P}_{y} &=&  - g\left(p_{i}^{2}\right) p_{yi} 
- a\, y_{i} + (2\, D)^{1/2} \xi_{i}(t)  
  \label{2d-stoch}
\end{eqnarray}
% \begin{equation}
%   \begin{array}{rcl}
%     \dot{x}_{i} =  p_{xi} - P_{x}  & \;;\;\; &
%     \dot{p}_{xi} - \dot{P}_{x} =  - g\left(p_{i}^{2}\right) p_{xi} 
% - a\, x_{i} + (2\, D)^{1/2}    \xi_{i}(t) \nonumber\\      
%     \dot{y}_{i} =  p_{yi} - P_{y} & \;;\;\; &       
%     \dot{p}_{yi} -\dot{P}_{y} =  - g\left(p_{i}^{2}\right) p_{yi} 
% - a\, y_{i} + (2\, D)^{1/2} \xi_{i}(t)  
%   \end{array}
%   \label{2d-stoch}
% \end{equation}
For $\bbox{P}=0$, i.e. for the initial conditions $p_{i}^{2}<p_{0}^{2}$
and sufficiently long times, this dynamics is equivalent to the motion of
free (or uncoupled) particles in a parabolic potential $U(x,y)=
a(x^{2}+y^{2})/2$ with the origin $\{0,0\}$. Thus, within this
approximation the system becomes a canonical-dissipative system again, in
the strict sense used in \sect{2}.  

\pic{fs-swarm} presents computer simulations of \eqn{2d-stoch} for the
relative motion of the particle swarm in the parabolic potential.
\footnote{The reader is invited to view a \emph{movie} of the respective
  computer simulations (t=0--130), which can be found at
  \texttt{http://ais.gmd.de/$^{\sim}$frank/swarm1.html} }
(Note that in this case all particles started from the same position
slightly \emph{outside} the origin of the parabolic potential. This has
been chosen in order to make the evolution of the different branches more
visible.)
As the snapshots of the spatial dispersion of the swarm show, we find
after an inital stage the occurence of two branches of the swarm which
results from a \emph{spontaneous symmetry break} (cf.  \pic{fs-swarm}
top).  These two branches will after a sufficiently long time move on two
limit cycles (as already indicated in \pic{fs-swarm} bottom). One of
these limit cycles refers to the left-handed, the other one to the
right-handed direction of motion in the 2d-space.  
\begin{figure}[htbp]
\centerline{\psfig{figure=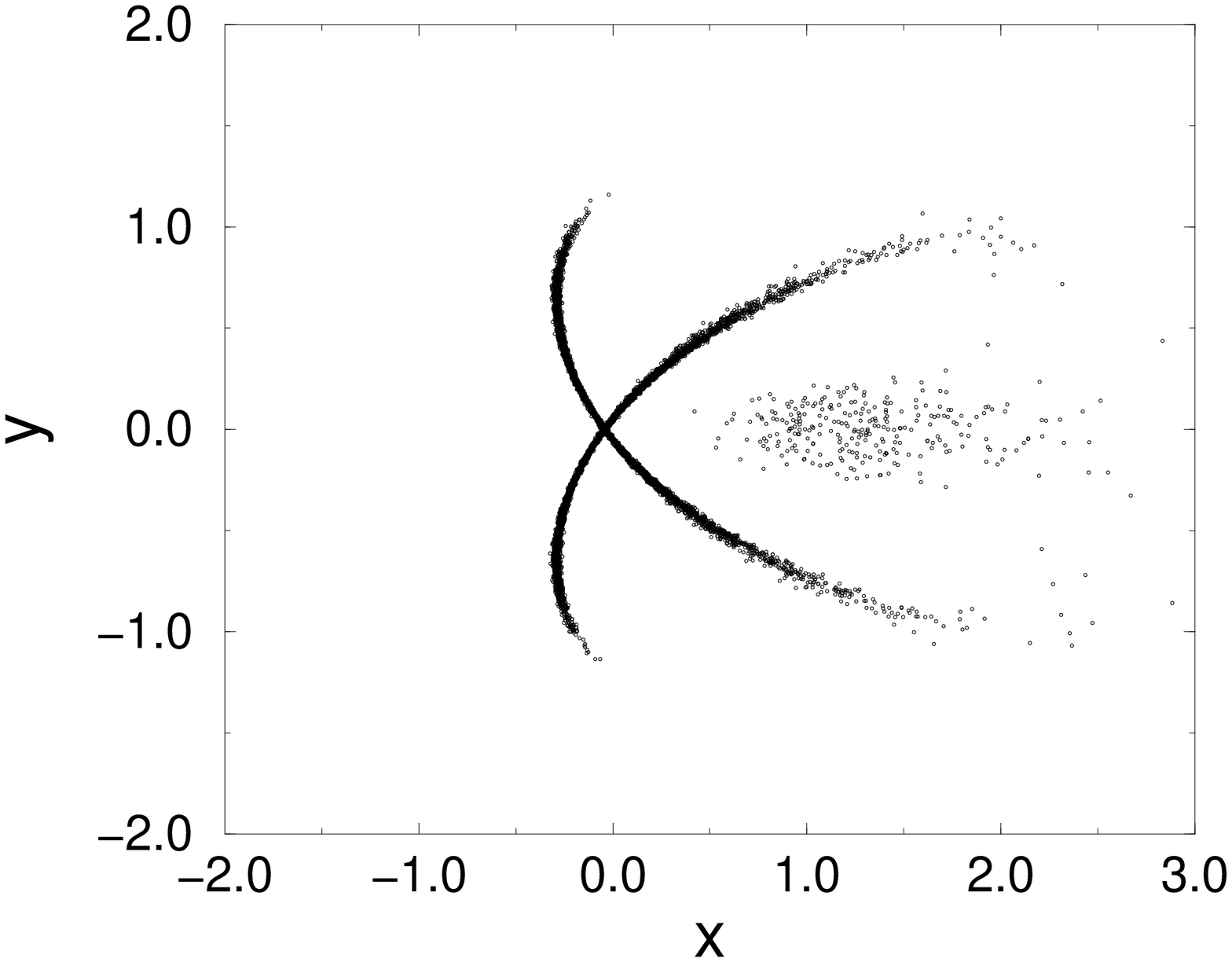,width=7.5cm}}

\centerline{\psfig{figure=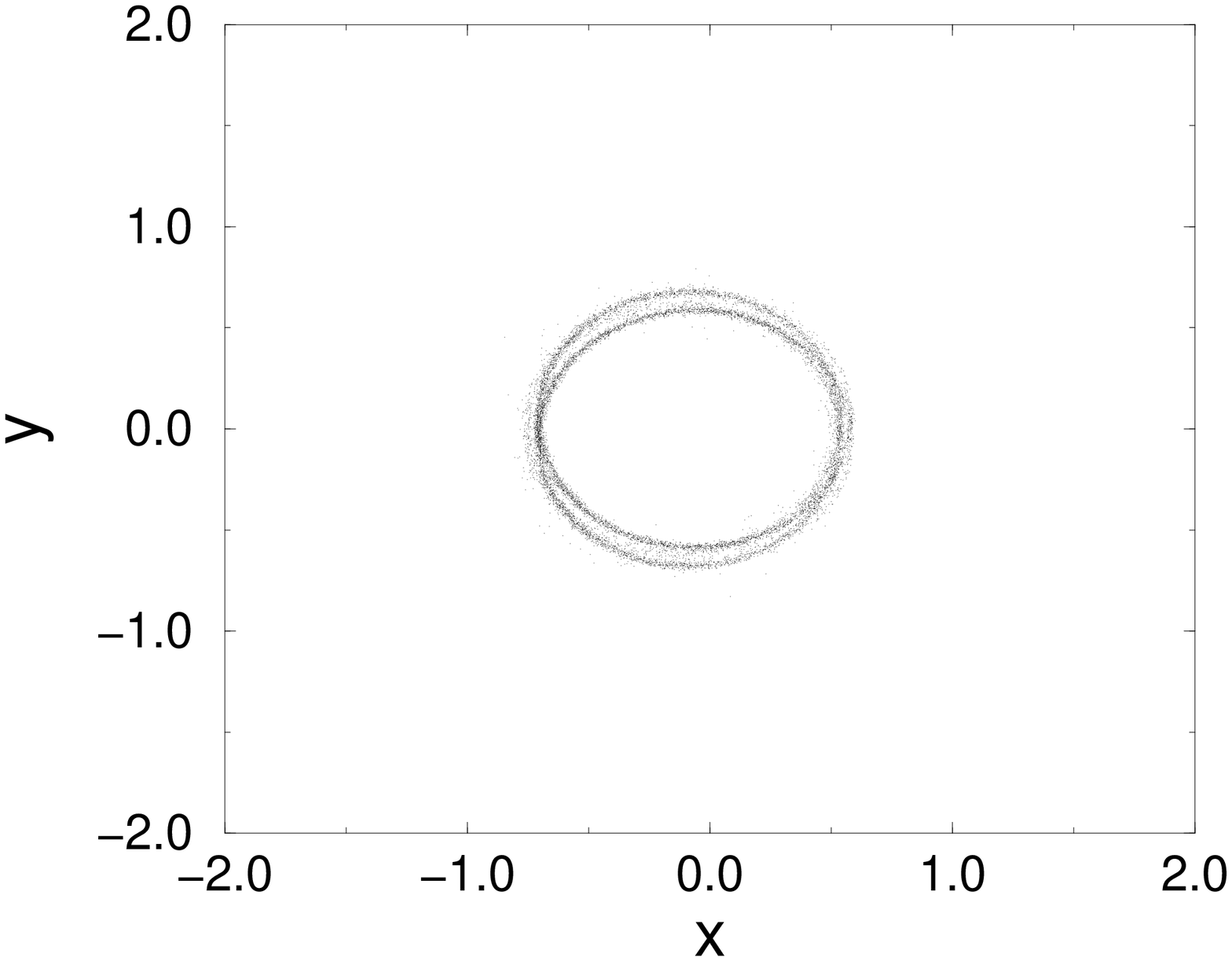,width=7.5cm}}
\caption[]{\label{fs-swarm}
  Snapshots (relative coordinates) of a swarm of $10.000$ particles
  moving according to \eqn{2d-stoch} with $\bbox{P}=0$. (top) $t=15$,
  (bottom) $t=99$.  Initial conditions: $\{x_{i},y_{i}\}=\{0.5,0.0\}$,
  $\{p_{xi},p_{yi}\}=\{0.0,0.0\}$ for all particles.  Parameters: $a=1$,
  $D=10^{-5}$, $s_0=10.0$; $c=1.0$; $\gamma_{0}=20$, $d_{2}=10$.
}
\end{figure}
This finding also agrees with the the theoretical investigations of the
deterministic case \cite{eb-fs-tilch-99} which showed the existence of a limit
cycle with the amplitude
\begin{equation}
r_0  = \abs{p_0}\,a^{-1/2}
\label{amplitude}
\end{equation}
provided the relation $s_{0}d_{2}>\gamma_{0}c$ is fulfilled. In the small
noise limit the radius of the limit cycles shown in \pic{fs-swarm} agrees
with the value of $r_{0}$. Further, \pic{vv-par8} has shown that the
averaged squared velocity $p^{2}(t)$ of the swarm indeed approaches the
theoretical value of \eqn{v-0}.

The existence of two opposite rotational directions of the swarm can be
also clearly seen from the distribution of the angular momenta $L_{i}$ of
the particles. \pic{l-99} shows the existence of a bimodal distribution
for $\rho(L)$. (The observant reader may notice that each of these peaks
actually consists of two subpeaks resulting from the initial conditions,
which are still not forgotten at $t=99$).  Each of the main peaks is
centered around the theoretical value
\begin{equation}
  \label{l0}
\abs{\bbox{L}}=L_{0}=r_{0}\,p_{0}   
\end{equation}
where $r_{0}$ is given by \eqn{amplitude} and $p_{0}$ is given by
\eqn{v-0}. 
\begin{figure}[htbp]
\centerline{\psfig{figure=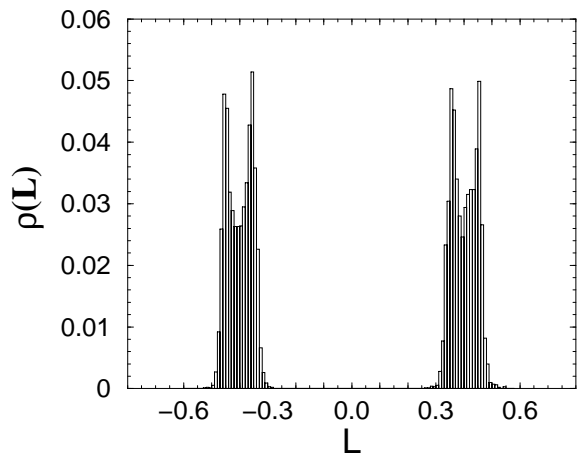,width=7.5cm}}
\caption[]{\label{l-99}
  Angular momentum distribution $\rho(L)$ for a swarm of $10.000$
  particles at $t=99$. The figure refers to the spatial snapshot of the
  swarm shown in \pic{fs-swarm} (bottom).}
\end{figure}

The emergence of the two limit cycles means that the dispersion of the
swarm is kept within certain spatial boundaries. This occurs after
a transient time used to establish the correlation between the individual
particles. In the same manner as the motion of the particles becomes
correlated, the motion of the center of mass is slowed down until it
comes to rest, as already shown in \pic{vv-par8}.

This however is not the case if the initial conditions $p_{i}^{2}(0)>
p_{0}^{2}$ are chosen. Then, the energy provided by the pumping does not
go completely into the relative motion of the particles and the
establishment of the limit cycles as discussed above. Instead, the center
of mass keeps moving as shown in \pic{rr-par8}, while the swarm itself
does not establish an internal order. \pic{snap-rr2} displays a snapshot of
the relative positions of the particles in this case (note the different
scales of the axes compared to \pic{fs-swarm}). 
\begin{figure}[htbp]
\centerline{\psfig{figure=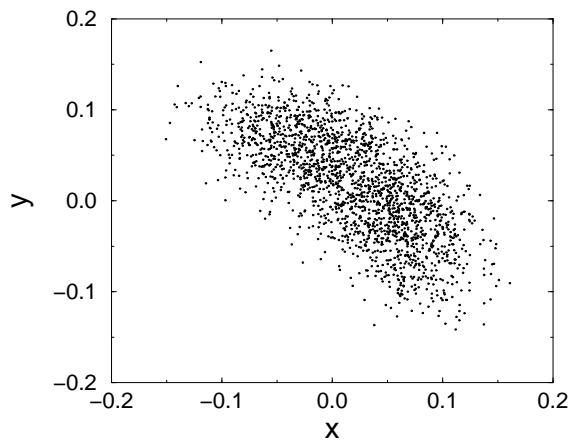,width=7.5cm}}
\caption[]{\label{snap-rr2}
  Snapshot (relative coordinates) of a swarm of $10.000$ particles moving
  according to \eqn{2d-stoch} at $t=99$. Initial conditions:
  $\{x_{i},y_{i}\}=\{0.5,0.0\}$, $\{p_{xi},p_{yi}\}=\{1.0,1.0\}$ for all
  particles.  Parameters see \pic{fs-swarm}.
}
\end{figure}

%%%%%%%%%%%%%%%%%%%%%%%%%%%
\subsection{Coupling via Mean Momentum and Mean Angular Momentum}
\label{4.2}
 
In the following we want to discuss two other ways of
\emph{global} coupling of the swarm which fit into the general framework
of canonical-dissipative systems outlined in \sect{2}. There, we have
introduced a dissipative potential
$G(I_{0},\bbox{I}_{1},\bbox{I}_{2},...)$ which depends on the different
invariants of motion, $\bbox{I}_{i}$. So far, we have only considered
$I_{0}=H$, \eqn{Hamil0} in the swarm model. If we additionally include
the mean momentum $\bbox{I}_{1}=\bbox{P}$, \eqn{mean-p} as the first
invariant of motion, the dissipative potential may read:
\begin{eqnarray}
\label{g-1}
G(I_{0},\bbox{I}_{1}) &=& \sum_{i=1}^{N} G_0(p_{i}^2) + G_{1}(\bbox{P}) \\
\label{g-11}
G_{1}(\bbox{P})&=& 
 \frac{C_{P}}{2} \left( \sum_{i=1}^{N} \bbox{p}_i - N\,\bbox{P_1} \right)^2
\end{eqnarray}
Here, $G_{0}(p^{2})$ is given by \eqn{g-0}. 
The stationary solution of the probability distribution,
$\rho^{0}(\bbox{p})$ is again given by \eqn{fpe-0-gen-inv}.
In the absence of an external potential $U(\bbox{r})$, the
\name{Langevin} equation that corresponds to the dissipation potential of
\eqn{g-1} reads now: 
\begin{eqnarray}
  \label{langev-p1}
\dot{\bbox{r}_{i}}&=&\bbox{p}_{i} \nonumber \\
\dot{\bbox{p}}_{i} &=& -g(p_{i}^{2})\, \bbox{p}_{i} - 
C_{P} \left(\sum \bbox{p}_i - N\,\bbox{P_1} \right) 
\nonumber \\
& & + (2\, D)^{1/2} \bbox{\xi}_{i}(t)
\end{eqnarray}
The term $G_{1}(\bbox{P})$ is choosen that way that it may drive the
system towards the prescribed momentum $\bbox P_1$, where the relaxation
time is proportional to $C_{P}^{-1}$.  If we would have a vanishing
dissipation function, i.e. $g=0$ for $p_{i}^{2}=p_{0}^{2}$, it
follows from \eqn{langev-p1} for the mean momentum:
\begin{equation}
  \label{pt}
  \bbox{P}(t)= \frac{1}{N}\sum_{i=1}^{N} \bbox{p}_{i}(t) = 
\bbox{P}_{1} + \Big[\bbox{P}(0)-\bbox{P}_{1} \Big] e^{-C_{p}t}
\end{equation}
The existence of two terms $G_{0}$ and $G_{1}$ however could lead to
competing influences of the resulting forces, and a more complex dynamics
of the swarm results. As before, this may also depend on the initial
conditions, i.e. $P^{2}_{1} \geq p_{0}^{2}$ or $P^{2}_{1} \leq
p_{0}^{2}$.

\pic{P_p_0.34} shows the squared velocity of the center of mass,
$P^{2}(t)$ and the averaged squared velocity of the swarm $p^{2}(t)$ for
 $P^{2}_{1} \leq
p_{0}^{2}$, We find an intermediate stage, where both velocities are
equal, before the global coupling drives the mean momentum $\bbox{P}$
towards the prescribed value $\bbox{P}_{1}$, i.e. $P^{2}(t)\to
(P_{1x}^{2}+P_{1y}^{2})$. On the other hand, $p^{2}(t)\to p_{0}^{2}$, as
we have found before for the force-free case and for the linearly coupled
case for similar initial conditions. The noticeable decrease of $P^{2}$
after the initial time lag can be best understood by looking at the
spatial snapshots of the swarm provided in \pic{snap-cp}. For $t=10$, we
find a rather compact swarm where all particles move into the same
(prescribed) direction. For $t=50$, the correlations between the
particles have already become effective, which means the swarm begins to
establish a circular front, which however does not become a full circle 
\footnote{The movie that shows the
  respective computer simulations for t=0--100 can be found at
  \texttt{http://ais.gmd.de/$^{\sim}$frank/swarm2.html}}. 
Eventually, we find again that the energy provided by the pumping goes
into the motion of the particles relative to the center of mass, while
the motion of the center of mass itself is driven by the prescribed
momentum. 
\begin{figure}[htbp]
\centerline{\psfig{figure=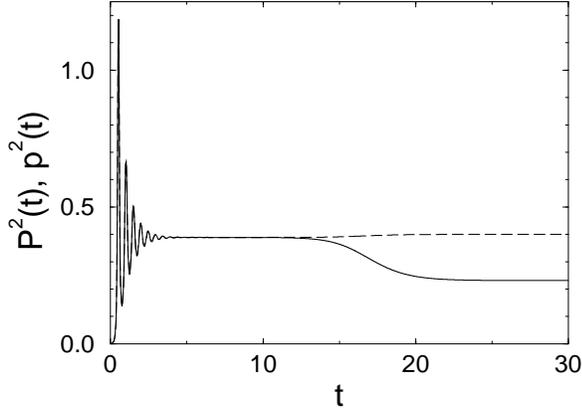,width=7.5cm}}
\caption[]{\label{P_p_0.34}
  Squared velocity of the center of mass, $P^{2}(t)=\Big(N^{-1}
  \sum_{i}\bbox{p}_{i}(t)\Big)^{2}$ (solid lines) and averaged squared
  velocity $p^{2}(t)=N^{-1}\sum p_{i}^{2}(t)$ (dashed lines) for the
  simulations shown in \pic{snap-cp}.}
\end{figure}
\begin{figure}[htbp]
\centerline{\psfig{figure=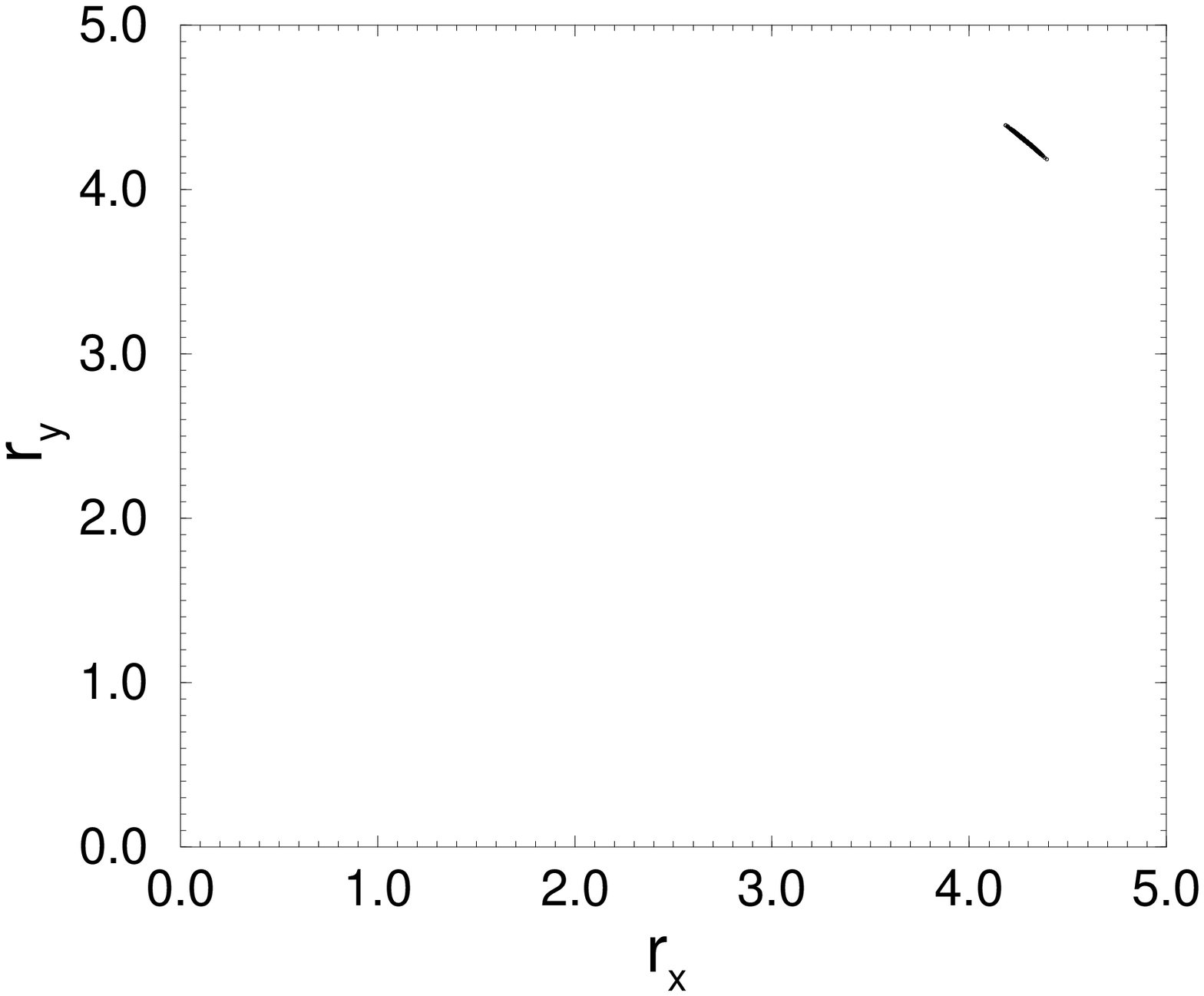,width=7.5cm}}

\centerline{\psfig{figure=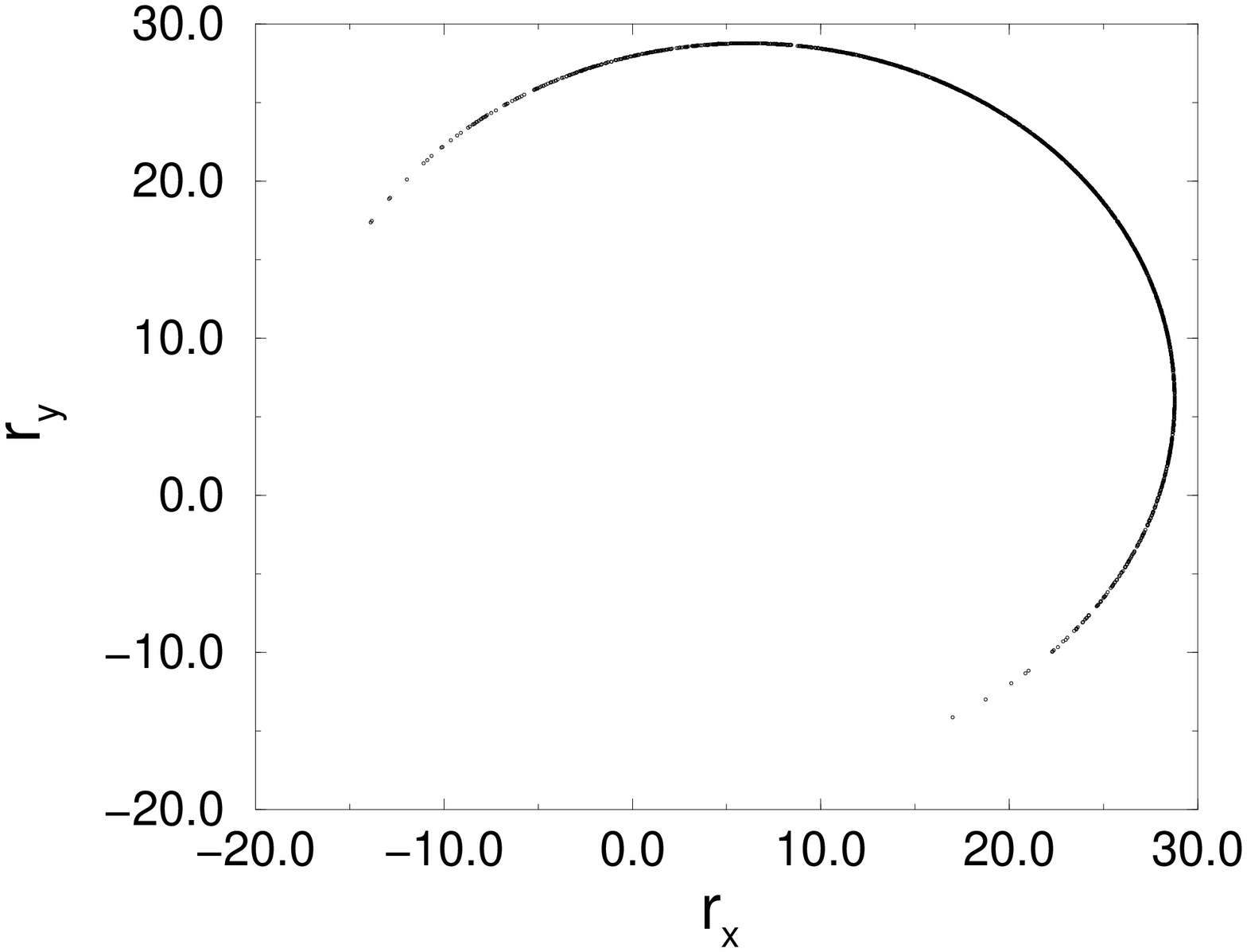,width=7.5cm}}
\caption[]{\label{snap-cp}
  Snapshots of a swarm of $2.000$ particles
  moving according to \eqn{langev-p1}. (top) $t=10$,
  (bottom) $t=50$.  Initial conditions: $\{r_{xi},r_{yi}\}=\{0.0,0.0\}$,
  $\{p_{xi},p_{yi}\}=\{0.0,0.0\}$ for all particles.  Parameters:
  $\{P_{1x},P_{1y}\}=\{0.344,0.344\}$, $C_{p}=10^{-3}$
  $D=10^{-8}$, $s_0=10.0$; $c=1.0$; $\gamma_{0}=20$, $d_{2}=10$.
}
\end{figure}

For the initial condition $P^{2}_{1} \geq p_{0}^{2}$ the situation is
different again, as \pic{P-p-1.0} shows. Apparently, both curves are the
same for a rather small noise intensity, i.e.  $P^{2}(t)=p^{2}(t)$ are
both equal, but different from $p_{0}^{2}$, \eqn{v-0} and the prescribed
momentum $P^{2}_{1}$. This can be only realized if all particles move in
parallel into the same direction. Thus, a snapshot of the swarm would
much look like the top part of \pic{snap-cp}.
\begin{figure}[htbp]
\centerline{\psfig{figure=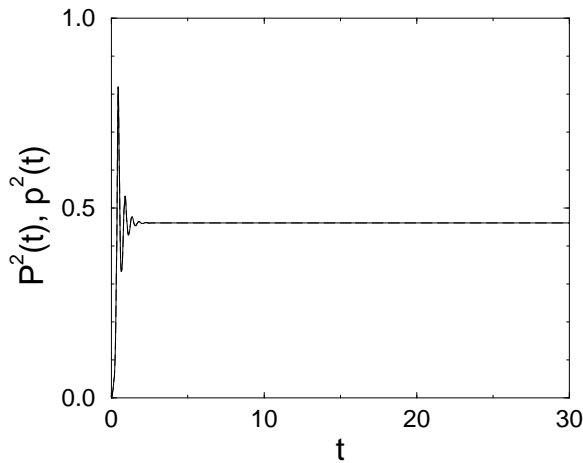,width=7.5cm}}
\caption[]{\label{P-p-1.0}
  Squared velocity of the center of mass, $P^{2}(t)=\Big(N^{-1}
  \sum_{i}\bbox{p}_{i}(t)\Big)^{2}$ (solid lines) and averaged squared
  velocity $p^{2}(t)=N^{-1}\sum p_{i}^{2}(t)$ (dashed lines) for a
  swarm of $2.000$ particles
  moving according to \eqn{langev-p1}. $\{P_{1x},P_{1y}\}=\{1.0,1.0\}$,
  for the other parameters and initial conditions see  \pic{snap-cp}.}
\end{figure}

Eventually, we may also use the second invariant of motion,
$\bbox{I}_{2}=\bbox{L}$, \eqn{mean-l} for a global coupling of the swarm.
In this case, the dissipative potential may be defined as follows:
\begin{eqnarray}
\label{g-2}
G(I_{0},\bbox{I}_{2})&=& \sum_{i=1}^{N} G_0(p_{i}^2) + G_{2}(\bbox{L})  \\
\label{g-22}
G_{2}(\bbox{L}) &=& \frac{C_{L}}{2} \left( \sum \bbox{r}_i \times \bbox{p}_i - N\,\bbox{L_1} \right)^2
\end{eqnarray}
$G_{0}(p^{2})$ is again given by \eqn{g-0}. The term $ G_{2}(\bbox{L})$
shall drive the system to a prescribed angular momentum $\bbox L_1$
with a relaxation time proportional to  $C_{L}^{-1}$.

$\bbox L_1$ can be used to break the symmetry of the swarm towards a
prescribed rotational direction.  In \sect{4.1} we have observed the
spontaneous occurence of lefthand and righthand rotations of a swarm of
linearly coupled particles. Without an additional coupling, both
rotational directions are equally probable in the stationary limit.
Considering both the parabolic potential $U(\bbox{r},\bbox{R})$, \eqn{parab-int} and
the dissipative potential \eqn{g-2}, the corresponding
\name{Langevin} equation may read now:
\begin{eqnarray}
  \label{langev-p2}
\dot{\bbox{r}_{i}}&=&\bbox{p}_{i} \nonumber \\
\dot{\bbox{p}}_{i}&=& -g(p_{i}^{2})\, \bbox{p}_{i} 
- \frac{a}{N} \sum_{j=1}^{N}\left(\bbox{r}_{i}-\bbox{r}_{j}\right)
\nonumber \\ && +
C_{L}\, \bbox{r}_{i} \times \left(\sum \bbox{r}_i
  \times \bbox{p}_i  -
N\,\bbox{L_1} \right)  \nonumber \\ 
&& + (2\, D)^{1/2} \bbox{\xi}_{i}(t)
\end{eqnarray}
The computer simulations shown in \pic{l-99-2} clearly display an
unimodal distribution of the angular momenta $L_{i}$ of the particles,
which can be compared to \pic{l-99} without coupling to the angular
momentum. Consequently, we find in the long time limit only one limit
cycle corresponding to the movement of the swarm into the same rotational
direction. The radius $r_{0}$ of the limit cycle is again given by
\eqn{amplitude}. 
\begin{figure}[htbp]
\centerline{\psfig{figure=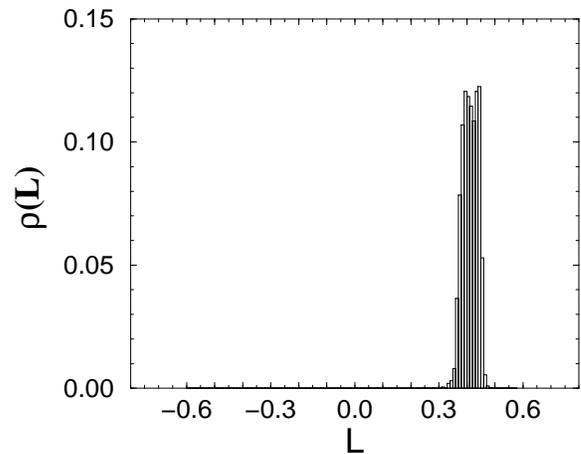,width=7.5cm}}
\caption[]{\label{l-99-2}
  Angular momentum distribution $\rho(L)$ for a swarm of $2.000$
  particles at $t=99$. For comparison with \pic{l-99}, the parameters,
  initial conditions and snapshot times are the same as in
  \pic{fs-swarm}. Additional coupling: $C_{L}=0.05,$,
  $\{L_{1z}\}=\{0.4\}$.}
\end{figure}

We would like to add that also in this case the dynamics depends on the
initial condition, $\bbox{L}_{1}$. For simplicity, we have assumed here
$\abs{\bbox{L}_1} = L_{0}= r_0 p_0$, \eqn{l0} which is also reached by
the mean angular momentum $\bbox{L}$ in the course of time (cf.
\pic{l-99-2}). For initial conditions $\abs{\bbox{L_{1}}}\ll L_{0}$,
there is of course no need for the rotation of \emph{all} particles into
the same direction. Hence we observe both left- and righthanded rotations
of the particles with different shares, so that the mean angular momentum
is still $\bbox{L}\to \bbox{L}_{1}$. This results in a broader
distribution of the angular momenta of the particles instead of the clear
unimodal distribution shown in \pic{l-99-2}. For initial conditions
$\abs{\bbox{L_{1}}}\gg L_{0}$ on the other hand, the stable rotation of
the swarm breaks down after some time, since the driving force
$\bbox{L}\to \bbox{L}_{1}$ tends to destabilize the attractor
$\bbox{L}\to \bbox{L}_{0}$. This effect will be investigated in a
forthcoming paper, together with some combined effects of the different
global couplings.

%%%%%%%%%%%%%%%%%%%%%%%%%%%%%%%%%%%%%%%%%%%%%%%%%%%%%
\section{Discussion}
\label{5}

Eventually, we can also combine the different global couplings discussed
above by defining the dissipation potential as:
\begin{eqnarray}
\label{g-12}
G(I_{0},\bbox{I}_{1},\bbox{I}_{2}) &=& G(p^2,\bbox{P},\bbox{L}) 
\nonumber \\ &=&
G_{0}(p^{2}) +G_{1}(\bbox{P})+ G_{2}(\bbox{L}) 
\end{eqnarray}
$G_{0}(p^{2})$ is given by \eqn{g-0}, $G_{1}(\bbox{P})$ by \eqn{g-11} and
$G_{2}(\bbox{L})$ by \eqn{g-22}. Considering further an additional --
external or interaction -- potential, 
the corresponding
\name{Langevin} equation can be written in the more general form: 
\begin{eqnarray}
  \label{langev-allg}
\dot{\bbox{r}_{i}}&=&\bbox{p}_{i} \nonumber \\
\dot{\bbox{p}}_{i}&=&  - g(p_i^2) \bbox{p}_i
- 
\frac{\partial}{\partial \bbox{r}_{i}} \left[ \alpha_0 \; U
  \left(\bbox{r},\bbox{R}\right)\right] 
\nonumber \\ && - \frac{\partial}{\partial \bbox{p}_i} 
\left[\alpha_1 \left(\bbox{P} - \bbox{P_{1}}\right)^2  +  
\alpha_2 \left(\bbox{L} - \bbox{L_{1}}\right)^2 \right]  
\nonumber \\ && + (2 D)^{1/2}\, \bbox{\xi}_i (t) 
\end{eqnarray}
The mean momentum $\bbox{P}$ and mean angular momentum $\bbox{L}$ are
given by \eqs{mean-p}{mean-l}; whereas the constant vectors
$\bbox{P}_{1}$ and $\bbox{L}_{1}$ are used to break the spatial or
rotational symmetry of the motion toward a preferred direction.  The
different constants $\alpha_{i}$ may decide whether the respective
influence of the conservative or dissipative potential is effective or not,
they further determine the time scale when the global coupling becomes
effective. The term $g(p^{2})$, \eqn{gh-qdv} on the other hand considers
the energetic conditions for the active motion of the swarm, i.e. it
determines whether the particle of the swarm are able to ``take off'' at
all.

The combination of the different types of coupling may lead to a rather
complex swarm dynamics, as already indicated in the examples discussed in
this paper. In particular, we note that the different terms may have
competing influences on the swarm, which then would lead to a
``frustrated'' dynamics with many possible attractors. 

In this paper, we have basically restricted the investigation of the
swarm dynamics to global couplings which fit into (or can be reduced
within some approximations to) the general outline of
canonical-dissipative systems.  Finally, we want to add some comments on
that. On one hand, it is possible to extend this kind of approach also to
other invariants of motion, this way e.g. covering previous
investigations of swarms coupled via the mean orientation of the
particles \citep{czirok-et-96,czirok-vicsek-00}. On the other hand, we
want to underline that canonical-dissipative systems are a theoretical
class of models, where \emph{both} conservative and dissipative elements
of the dynamics are determined by invariants of the mechanical motion.
Thus, from this perspective, a more realistic swarm dynamics may be also
based on less restrictive assumptions.

The advantage to use canonical-dissipative systems as a framework for
swarm dynamics is given by the fact that in many cases the rather complex
dynamics can be mapped to an analytically tractable model.  With the
Hamiltonian theory of many-particle systems as a starting point, we are
able to extend known solutions for conservative systems to
non-conservative systems. This allows us to to construct a
canonical-dissipative system, the solutions of which converges to the
solution of the conservative system with given energy. That means, for
given initial conditions, we could predict the asymptotic solution of the
canonical-dissipative dynamics by means of the solutions of the
Hamiltonian equations on the respective energy surface.

In addition to these theoretical considerations which have a value of
their own, we want to note that the framework of canonical-dissipative
systems still covers important features of real (biological) systems,
such as energy take-up and dissipation.  The general description outlined
in this paper allows us to gradually add more and more complexity to the
swarm model, this way bridging the gap between a known (physical)
dynamics to a more complex (biological) dynamics
\citep{eb-fs-01-pas,eb-fs-01}. Some hints for this shall be given at the
end.

On the level of the ``individual'' particles, the key dynamics of the
model is given by a modified Langevin equation that, in addition to
stochastic influences, also considers other forces on the particle,
resulting e.g. from external potentials, interactions, use of stored
energy \citep{eb-fs-tilch-99}, or influences of a self-consistent field 
\citep{lsg-fs-mieth-97,lsg-mieth-rose-malch-95} that already exceed the
framework of canonical-dissipative systems. 

The consideration of an \emph{external} potential also allows to model
the \emph{spatial environment} of the swarm, for instance to consider
\emph{obstacles} \citep{fs-eb-tilch-98-let}. Additionally, we can also
consider that the pumping of energy for the particles is restricted to
certain spatial domains which model \emph{food sources}. In this case the
dissipation function $g(H)$ also becomes a spatial function. Such an
extension has been already discussed in \citep{steuern-et-94,
  fs-eb-tilch-98-let} and can be also implemented in the swarm model
discussed here.  Eventually, we note that the genuin particle-based
approach to collective phenomena used in this paper is not restricted to
biological systems, but also applicable for describing and simulating
complex interactive systems in a wide range of applications, even in
economics and social systems \citep{fs-book-01}.

%%%%%%%%%%%%%%%%%%%%%%%%%%%%%%%%%%%%%%%%%%%%%%%%%%%%%%%%%%%%%%

\begin{acknowledgments}
The authors thank J. Dunkel and U. Erdmann for discussions.
\end{acknowledgments}

%\bibliography{your bib file}
%\bibliography{swarm,ego-new,pumped-new,bio-new,habil}
\bibliography{fs-eb-t-pre01}
\end{document}